\documentclass{aastex631}
\usepackage[utf8]{inputenc}

\shorttitle{Study of the Soft X-ray Emission Lines in NGC 4151}
\shortauthors{Grafton-Waters et al.}

\begin{document}

\title{A Study of the Soft X-ray Emission Lines in NGC 4151 \\
I. Kinematic Properties of the Plasma Wind}
%\maketitle

%\correspondingauthor{S. Grafton-Waters}
%\email{sam.waters.17@ucl.ac.uk}
\author[0000-0002-4833-8612]{S. Grafton-Waters}
\affiliation{Mullard Space Science Laboratory, University College London, Holmbury St. Mary, Dorking, Surrey RH5 6NT, UK}
%\email{sam.waters.17@ucl.ac.uk}

\author{M. Ahmed}
\author{S. Henson}
\author{F. Hinds-Williams}
\author{B Ivanova}
\author{E. Marshall}
\author{H. Udueni}
\author{D. Theodorakis}

\affiliation{Nottingham University Academy of Science and Technology, 93 Abbey St, Lenton, Nottingham NG7 2PL}

\author[0000-0002-0383-6917]{W. Dunn}
\affiliation{Mullard Space Science Laboratory, University College London, Holmbury St. Mary, Dorking, Surrey RH5 6NT, UK}

\begin{abstract}
We present our analysis of the narrow emission lines produced in the plasma regions within the bright active galactic nucleus of NGC 4151, from an ORBYTS research-with-schools public engagement project. Our goal was to test whether the properties of these plasma regions changed between XMM-Newton observations spanning 15 years from 2000 to 2015, by measuring the outflow velocities and distances. From this study, we found that NGC 4151 has at least two to three plasma regions. There is no evidence of the outflowing wind properties changing as the velocities and distances are consistent throughout the observations.  %\textbf{(95 words)}
\end{abstract}

\section{Introduction}

\begin{figure}
	\centering
	\epsscale{1.1}
	\plotone{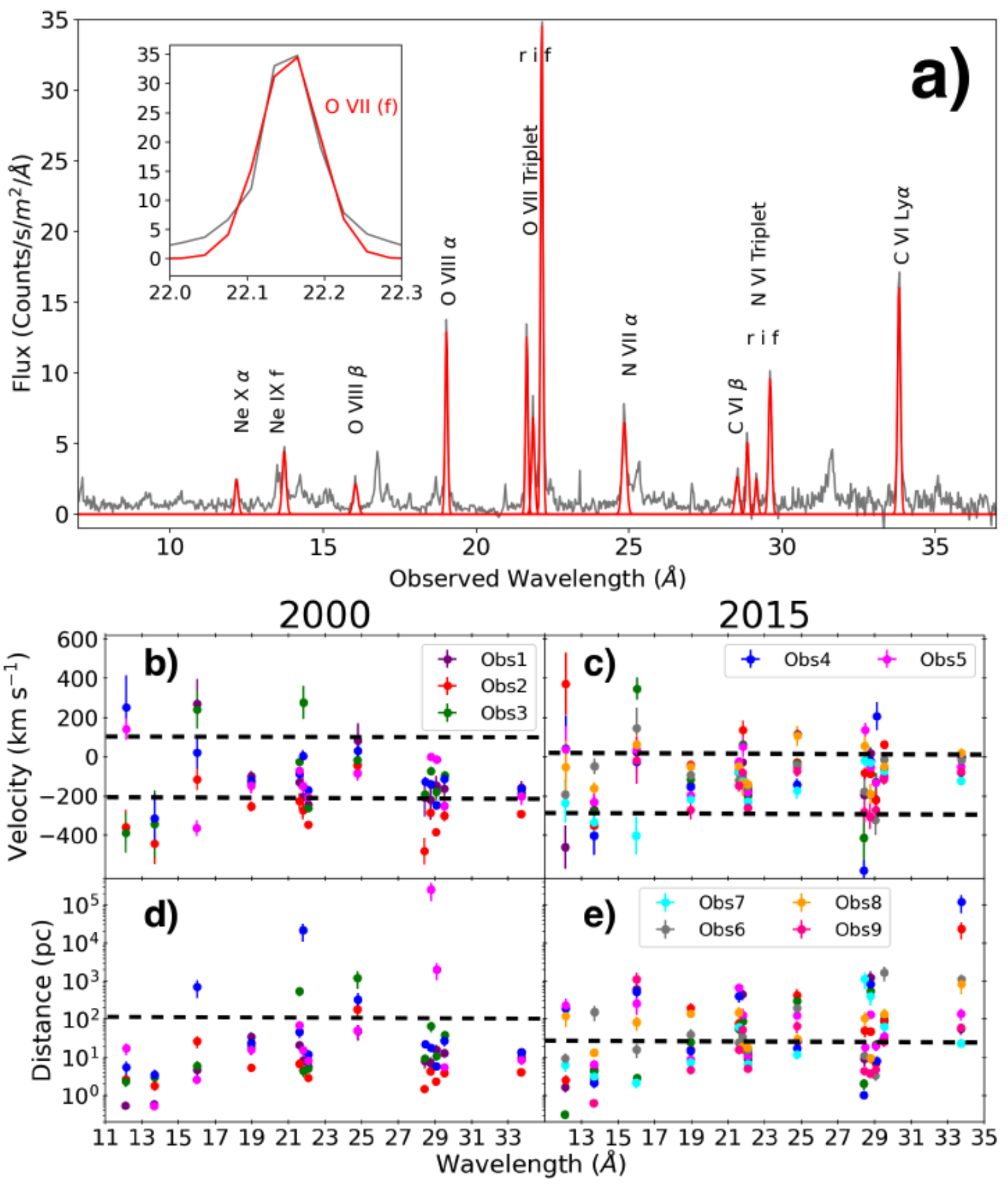}
	\caption{\textbf{(a)} RGS spectrum (grey line) with the Gaussian models (red lines) fitted to the strongest emission lines. The lines we investigated are labelled with their respective ions. The r, i and f labels stand for resonance, intercombination and forbidden, respectively, corresponding to electron transitions in the He-like triplet ions \citep[see e.g.][for details]{Porquet2000}. The insert shows the OVII f line fitted with its respective Gaussian. \textbf{(b-c)} Measured line velocities using Eq. \ref{EQ_Line_Vel} and \textbf{(d-e)} calculated distances from Eq. \ref{EQ_Distance} are compared for 2000 (left) and 2015 (right). Colours indicate different observations from each epoch, as labelled in the legends, and the black dashed lines signify the boundaries between plasma regions. } %\textbf{109}
	\label{Fig:Figure}
\end{figure}

NGC 4151 (z = 0.003262) is a type 1.5 Seyfert active galactic nucleus (AGN) with a black hole (BH) mass of $M_{BH} = 3.59 \times 10^7$ M\textsubscript{$\odot$} \citep{Bentz_Katz2015}. While NGC 4151 is highly variable in the hard X-ray band \citep[e.g.][]{Beuchert2017}, the soft X-ray spectrum is very similar to that of NGC 1068 with multiple strong emission features on top of an almost negligible continuum \citep[e.g.][]{Grafton-Waters2021}. %consistent over each observation \citep{Schurch_RGS, Beuchert2017}. 
This extreme contrast between soft and hard X-ray bands makes NGC 4151 an ideal target to study the properties of the narrow line region (NLR). The NLR is part of the outflowing wind, composed of separate plasma regions, that produces the emission lines observed in the X-ray spectra (see Figure \ref{Fig:Figure}). The aim of this ORBYTS research-with-schools public engagement was to assess whether the NLR properties changed between observations spanning a 15 year period, by measuring the outflow velocities and distances of each region. The code can be found at \cite{sgwxray_2021_5116838}. %\textbf{(161 words)}

\section{Method}

NGC 4151 was observed 27 times by XMM-Newton during six separate epochs between 2000 and 2015. Here we analysed the reflection grating spectrometer \citep[RGS;][]{denHerder2001} data from each observation. The data were reduced using the \texttt{RGSPROC} command in the \texttt{SAS} software \texttt{v 17.0.0}\footnote{See \url{https://www.cosmos.esa.int/web/xmm-newton/sas-thread-rgs}}. Any large background counts in CCD9 were removed before combining RGS1 and RGS2 data with \texttt{RGSCOMBINE}.

We modelled the thirteen strongest emission lines (labelled in Figure \ref{Fig:Figure}) that were present in all X-ray spectra with a simple Gaussian model in Python, defined as
\begin{equation}
G(x) = \frac{A}{\sqrt{2\pi \sigma^2}}\exp{\left[-\frac{(x - \mu)^2}{2\sigma^2}\right]},
\label{Eq_Gaus}
\end{equation}
where \textit{A} is the amplitude, $\sigma$ is the standard deviation, and $\mu$ is the mean value (the emission line centre). To measure the properties of each emission line we defined a wavelength range ($\lambda_R$) to ensure we were only fitting one line at a time. This is shown in the insert in Figure \ref{Fig:Figure} (a). This Gaussian model and the X-ray spectrum were then fed into the \texttt{LMFIT} interface in Python \citep{LMFIT2014}, which returned the parameter values ($\mu$, $\sigma$, and \textit{A}) and errors. 

To estimate the outflow velocity ($v_{out}$) of the plasma regions, we used the redshift ($z$) equation, given by
\begin{equation}
z = \frac{v_{out}}{c} = \frac{\lambda_{obs} - \lambda_{rest}}{\lambda_{rest}},
\label{EQ_Line_Vel}
\end{equation}
where \textit{c} is the speed of light, and $\lambda_{obs}$ and $\lambda_{rest}$ are the observed and rest frame wavelengths, respectively. The velocity shift was measured for each emission line individually. The observed wavelength ($\mu$) was measured from our modelling and the rest wavelengths were obtained from the \texttt{SPEX} line list\footnote{\url{https://personal.sron.nl/~kaastra/leiden2020/line_new.pdf}}. We note that it is not the emission line that has the velocity, but instead it is a property of the line-emitting plasma region. If many of the emission lines had similar values they were likely to originate from the same plasma region within the outflowing wind. To estimate the distances ($R$) from the BH we assumed that $v_{out}$ was greater than or equal to the escape velocity ($v_{esc}$) of the BH, given by
\begin{equation}
R = \frac{2GM_{BH}}{v^2_{esc}},
\label{EQ_Distance}
\end{equation}
where \textit{G} is the gravitational constant, and $M_{BH}$ is the BH mass.

After modelling each emission line, and obtaining the velocities and distances from each observation, we compared results. This allowed us to evaluate whether the wind properties changed over time. %\textbf{(351 words)}

\section{Results}
Figure \ref{Fig:Figure} (a) shows the 2000 RGS spectrum with the Gaussian models fitted to the strongest emission lines; these lines are labelled. The insert in Figure \ref{Fig:Figure} (a) displays the OVII forbidden line at 22.15 \AA\ with the Gaussian model fitted on top. From the spectral modelling of all the observations, we compare the results for the velocities and distances in 2000 (left) and 2015 (right) in Figure \ref{Fig:Figure} (b-c) and (d-e), respectively. The black dashed lines in Figure \ref{Fig:Figure} indicate the plasma boundaries for $v_{out}$ and $R$, suggesting that there are at least two to three plasma regions. However, we cannot rule out the possibility of more. The results for 2000 and 2015 are consistent with the velocities and distance from all observations. %- i.e. multiple lines with similar $v_{out}$ and $R$ values that originate from the same region. Our analysis suggests that there are at least two to three plasma regions, 

Our aim was to see whether the outflowing wind properties changed over the course of 15 years. However, based on this investigation, we found no evidence of this. One may expect that from 2000 to 2015, the distances and velocities would have increased and decreased, respectively, as the plasma regions move away from the BH. However, assuming the plasma is travelling at 500 km s\textsuperscript{-1}, the distance travelled in a 15 year time period would only be 0.008 pc, which is insignificant compared to the distances that we measured (i.e. $> 1$ pc). One explanation could be that we are viewing the plasma perpendicular to the outflow direction, meaning we are unable to observe the true outflow velocity, since the Doppler shifting of lines would only represent a component of its true outflow velocity. Alternatively, the plasma has been relatively static over the 15 years, which is reasonable given the large distances from the BH. In that case, 15 years may simply not be enough time for us to observe any significant changes with regards to the outflowing wind in NGC 4151. 

We did have some problems when modelling the data, which could have affected the final results. For example, the line centre ($\mu$) depended on $\lambda_R$ which we set when modelling the lines. This meant that if $\lambda_R$ was too large, then multiple lines would be fitted together by the model, or if $\lambda_R$ was too small we would lose information about the feature. This was a particular problem if $\lambda_R$ for an emission line differed between observations, especially if the S/N ratios were poor. As a result, a slightly different velocity was obtained each time.

%This was a particular problem if the S/N ratio of one spectrum was poor such that an emission feature was weak compared to another observation. As a result, $\lambda_R$ would be different between the two  and therefore a slightly different velocity was obtained each time. 

%This likely explains why the distances, and thus velocities, are consistent over all observations.%This could explain why some lines have distances that span over four orders of magnitude between observations, e.g. 21.6 \AA\ and 33.6 \AA\ in Figures \ref{Fig:Figure} (d) and (e), respectively.

%\textbf{(457 words)}

\section{Conclusions}
From our RGS analysis on NGC 4151, we found that there are at least two to three plasma regions that produce the emission features in the outflowing wind. However, the velocities and distances are consistent throughout the observations. This implies that either the plasma regions have been relatively static over the 15 year observation period, or because we are seeing the outflowing wind perpendicular to the flow direction, we are not measuring the velocities in the direction of motion. %\textbf{84 words}

\begin{acknowledgments}
This work was undertaken with the ORBYTS Research-with-schools public engagement project, partnering scientists with schools to support students’ involvement in space research. S.G.W. acknowledges the support of a PhD studentship awarded by the UK Science \& Technology Facilities Council (STFC).  %\textbf{39 words}
\end{acknowledgments}

%\bibliography{references.bib}	
%\bibliographystyle{aasjournal}

\end{document}